# Modeling Hormesis Using a Non-Monotonic Copula Method

Farzaneh Ghasemi Tahrir[1,*]

**Abstract**

This paper presents a probabilistic method for capturing non-monotonic behavior under the biphasic dose-response regime observed in many biological systems experiencing different types of stress. The proposed method is based on the 'rolling-pin' method introduced earlier to estimate highly nonlinear and non-monotonic joint probability distributions from continuous domain data. We show that the proposed method outperforms the conventional parametric methods in terms of the error (namely RMSE) and it needs fewer parameters to be estimated *a priori*, while offering high flexibility. The application and performance of the proposed method are shown through an example.

**Keywords.** Hormesis, non-monotonicity, rolling-pin method, monotonization, copula joint probability

## 1. Introduction

Several decades after Southam and Ehrlich used the term *hormesis* for the first time to refer to the biphasic dose-response behavior in the context of toxicology [1], this type of behavior is being revisited in many biological systems undergoing different treatments [2]. Hormesis is characterized by the U-shaped (or inverted U-shaped) response of a system to different values of the stressor. This means the system shows different inhibitory or stimulatory behavior when exposed to low or high doses of the stressor. It has been reported that over 600 chemicals show biphasic dose response relationships [3, 4]. Hormesis, therefore, has become a frequently observed phenomenon in many popular fields of research in biology including mitochondria research and aging [5, 6, 7].

Modeling the hormetic behavior has been of interest in different fields as it can relate the stressor value to the observed output and provide a basis for assessing its potential risks. Some of the studies focused on the mechanistic description of the phenomenon, while some others approached the problem from the black-box modeling point of view. Among those, the log-logistic-based models [8, 9] and their different extensions [10, 11] are notable.

In this paper we introduce a dose-response modeling method based on the joint probability distributions for capturing the non-monotonicity of the hormetic systems. Different varieties of joint probabilities have been introduced to account for the non-monotonic behavior including the moment based multivariate probabilities [12, 13] and copula based methods [14, 15, 16]. In this work we adopt a copula-based joint probability estimation method called the "rolling-pin" method. The RP method combines a so-called monotonization transformation with the conventional parametric copula method [17], enabling the user to capture highly nonlinear and non-monotonic behavior underlying the data with a relatively low computational complexity. This way we are able to model the hormetic U-shaped behavior with minimum number of parameters, high flexibility and low computational cost. The multivariate nature of the RP method makes it possible to work with higher number of input (dose) and output (response) variables. The application and performance of the method is demonstrated using an example.

[1]: Department of Neuroscience, Lewis Katz School of Medicine, Temple University, Philadelphia, PA 19140
[*]: Corresponding Author: tue59234@temple.edu

The rest of the paper is organized as follows. In the method section we describe our expectation and maximum a posteriori methods of dose-response relationship determination based on the RP method. Section 3 is dedicated to an illustrative example. Finally, in section some concluding remarks are provided.

## 2. Method

### 2.1. Monotonization Transformation

As introduced by Mohseni Ahooyi et al. [14], the following set of monotonization transformation can be applied to monotonize a $d$-dimensional vector $\mathbf{X} = (X_1, \ldots, X_d)^T$ of continuous random variables with respect to a reference variable $X_r$ and a vector of monotonizing parameters $\boldsymbol{\alpha_m} = (\alpha_{1,m}, \ldots, \alpha_{d,m})^T, \alpha_{i,m} \in [0,1]$. Mohseni Ahooyi et al. showed that choosing an appropriate $\boldsymbol{\alpha_m}$ not only monotonizes the resulting random vector, $\mathbf{Y}$, components with respect to $X_r$, but also ensures the pairwise monotonic relationship between each $Y_i$ and $Y_j$, $i, j \in \{1, \ldots, d\}$. The monotonization transformation is defined as follows:

$$Y_i = (1 - \alpha_{i,m})X_i + \alpha_{i,m}X_r, \quad i \in \{1, \ldots, d\} \tag{1}$$

where $Y_i$ denotes the monotonized variable using the original variable $X_i$. $X_r$ can be chosen from $X_i$s or can be an independent continuous random variable. For more information regarding the selection criteria of $X_r$ and estimation of $\boldsymbol{\alpha_m}$ see [14]. It is recommended in that reference to first center and standardize $X_i$s.

### 2.2. Estimating the Joint Probability Distribution

As the monotonization transformation is a one-to-one transformation, the following relationship can be shown between the joint probability density functions of $\mathbf{X}$ and $\mathbf{Y}$, i.e. $f_\mathbf{X}(\mathbf{x})$ and $f_\mathbf{Y}(\mathbf{y})$:

$$f_\mathbf{X}(\mathbf{x}) = f_\mathbf{Y}(\mathbf{y})|\det(\mathbf{J})| \tag{2}$$

where $\mathbf{J}$ denotes the Jacobian matrix $\partial \mathbf{y} / \partial \mathbf{x}$. Calculating $|\det(\mathbf{J})|$ for the monotonization transformation, Eq.(2) becomes:

$$f_\mathbf{X}(\mathbf{x}) = f_\mathbf{Y}(\mathbf{y}) \prod_{i=1}^{d}(1 - \alpha_{i,m}) \tag{3}$$

As $Y_i$s are monotonized, a copula density function [17] can be used to estimate $f_\mathbf{Y}(\mathbf{y})$ and therefore $f_\mathbf{X}(\mathbf{x})$ can be estimated using a conventional parametric copula regardless of the nonlinearity or non-monotonicity of the relationships between each $X_i$ and $X_j$, $i, j \in \{1, \ldots, d\}$:

$$f_\mathbf{X}(\mathbf{x}) = c\left(F_{Y_1}(y_1), \ldots, F_{Y_d}(y_d)\right) \prod_{i=1}^{d}(1 - \alpha_{i,m})f_{Y_i}(y_i) = \frac{\partial C\left(F_{Y_1}(y_1), \ldots, F_{Y_d}(y_d)\right)}{\partial F_{Y_1}(y_1) \ldots \partial F_{Y_d}(y_d)} \prod_{i=1}^{d}(1 - \alpha_{i,m})f_{Y_i}(y_i) \tag{4}$$

where $c: [0,1]^d \to \mathbb{R}^+ \cup \{0\}$ is the copula density function, $f_{X_i}, f_{Y_i}: \mathbb{R} \to \mathbb{R}^+ \cup \{0\}$ denote the marginal density functions of $\mathbf{X}$ and $\mathbf{Y}$, $F_{Y_i}: \mathbb{R} \to [0,1]$ is the marginal cumulative distribution function of $Y_i$ and $C: [0,1]^d \to [0,1]$ is an appropriate parametric copula function, respectively [14]. The marginal probability densities can be estimated parametrically, non-parametrically or semi-parametrically.

## 2.3. Data-Based Estimation of Dose-Response Curves

The estimated multivariate density function $f_\mathbf{X}(\mathbf{x})$ can be conveniently utilized to find the arbitrary (non-monotonic) functions of multiple dose and response variables. Here we present this application for only one dose and one response variables. Once $f_{\mathbf{Dose,Response}}(Dose, Response)$ is estimated using an appropriate reference variable (possibly *Dose*), $\alpha$ and an appropriate parametric copula (Gaussian, Student's t, Frank, etc.), the functionality of the pair $(Dose, Response)$ can be calculated using either of these ways:

*I)*   *Expectation Method (Mean)*

In this method, the conditional probability of the response variable given the amount of *Dose* is calculated for values of *Dose* within its domain.

$$\overline{Response}(Dose) = \int_{\mathbf{R}_{Response}} (f(Response|Dose).Response)\, dResponse \qquad (5)$$

where $f(Response|Dose)$ denotes the conditional probability and $\mathbf{R}_{Response}$ is the range of response. $\overline{Response}(Dose)$ refers to the estimated functionality of the response with respect to the dose variable. The conditional probability of Eq. (5) can be estimated as:

$$f(Response|Dose) = \frac{f(Response, Dose)}{f(Dose)} = \frac{f(Response, Dose)}{\int (f(Response, Dose))\, dResponse} \qquad (6)$$

where $f(Response, Dose)$ denotes the joint distribution of the dose and response variables estimated through Eq. (4).

*II)*   *Maximum a posteriori (MAP) Method*

In this method the value of the response variable that gives rise to the maximum conditional probability of Eq.(6) is calculated for each query value of the dose variable.

$$\text{Response}_{MAP}(Dose) = \arg\max_{Response} \left(f(Response|Dose)\right) \qquad (7)$$

This method is particularly useful if the conditional probability of Eq.(6) is unimodal.

In the next section we provide an example of the method application in estimating the $(Dose, Response)$ relationship.

## 3. Application Example

This section is dedicated to demonstrate the application of the method of biphasic dose-response relationship estimation. To this end, a sample dataset of 30 data points was simulated where the logarithm of dose data distributed according to a normal distribution. The marginal densities of $\log(Dose)$ and $Response$ were then estimated non-parametrically using the Gaussian kernel density estimation [18]. $Response$ data were monotonized with respect to $\log(Dose)$ using 6 different values of $\alpha$: 0 (no monotonization), 0.1, 0.3, 0.5, 0.7 and 0.95. The corresponding rolling-pin joint probability densities of $\log(Dose)$ and $Response$ then were derived according to Eq. (4) using a Gaussian copula. This is an appropriate choice as the monotonization will make the dependence structure of the monotoniazed variables converge to that of $(\log(Dose), \log(Dose))^T$ which is close to Gaussian. The simulated dataset and the contour plots of the RP-estimated joint probabilities for different values of $\alpha$ are shown in Fig. 1. Associated with each joint probability, the expectation and MAP estimated dose-response curves were calculated according to Eqs. (5)-(7). The RMSE of each estimated curve was then calculated with respect to the dataset for the expectation ($R_{MEAN}$) and maximum a posteriori ($R_{MAP}$) methods, listed in Fig. 1. It can be seen that as $\alpha$ approaches 1, the RP method gives a better estimation of the U-shaped non-monotonic behavior presented by the data and the RMSE decreases, indicative a better estimation of the dose-response curve represented by the data. It should be noted that as $\alpha$ gets very close to 1, there can happen some information loss [14] and therefore a trade-off scheme should be applied to increase the accuracy while avoiding overfitting or information loss. It is also worth mentioning that in capturing the non-monotonic behavior in the data, minimal assumptions about the functionality of the dose and response were made (only the Gaussian copula), and by only tuning one parameter ($\alpha$) we were able to achieve higher complexity (adding non-monotonicity).

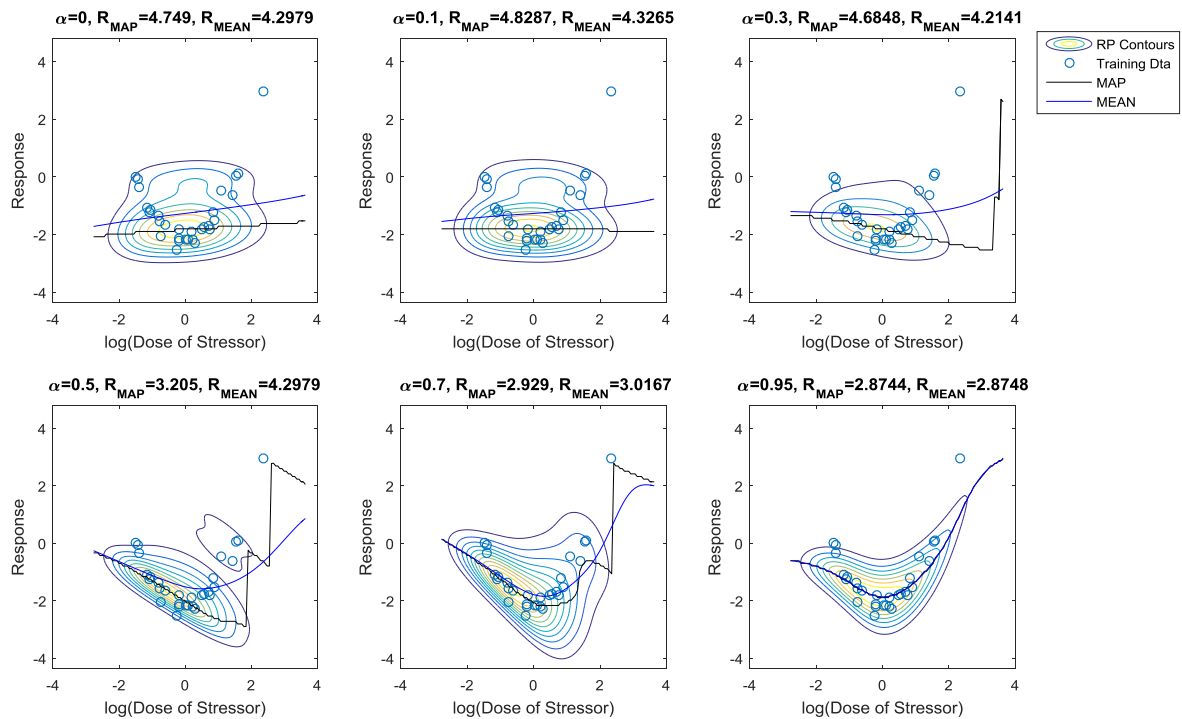

**Figure 1:** The mean- and MAP-estimated dose-response curves using the RP-method-estimated joint probabilities for different values of the monotonization parameter, $\alpha$.

## 4. Conclusions

In this work we employed a previously introduced rolling-pin method of joint probability estimation to model the biphasic dose-response relationship. As the RP method is capable of capturing highly nonlinear and non-monotonic relationships as well as dealing with complex and unknown dependence structures with minimum assumptions and parameters, it provides a useful framework in modeling the hormetic effect. We demonstrated the application of the method to an example and concluded that the biphasic behavior of the dose-response curve can be well-estimated with only a selection the Gaussian kernel for the marginal densities, Gaussian copula in the RP method and adjusting the monotoization parameter ($\alpha$). This shows that with adequately large values of $\alpha$ (closer to 1) higher degrees of nonlinearity and non-monotonicity can be modeled without any need to higher model complexities and parameters. In addition to this flexibility, the method is computationally tractable and is capable of dealing with higher numbers of input and output variables due to the multivariate nature of the copula-based RP method.